\documentclass{aa}
\usepackage{graphicx}
\usepackage{amssymb}

\usepackage{color}
\usepackage{natbib}
\usepackage{tabularx}
\usepackage{url}

\newcommand{\HESSJ}{HESS~J1943+213}
\newcommand{\MASS}{2MASS~J19435624+2118233}

\def\arcmin{\hbox{$^\prime$}}
\def\arcsec{\hbox{$^{\prime\prime}$}}

\begin{document}


\title{The host galaxy and \textit{Fermi}-LAT counterpart of \HESSJ}

\author{Peter, D.\inst{1}, Domainko, W.\inst{2}, Sanchez, D. A.\inst{3}, van der Wel, A.\inst{1} and G\"assler, W.\inst{1}}

\institute{Max-Planck-Institut f\"ur Astronomie, K\"onigstuhl 17, D-69117 Heidelberg, Germany
\and
Max-Planck-Institut f\"ur Kernphysik, Saupfercheckweg 1, D-69117 Heidelberg, Germany
\and
Laboratoire d'Annecy-le-Vieux de Physique des Particules, Universit\'{e} de Savoie, CNRS/IN2P3, F-74941 Annecy-le-Vieux, France}

\offprints{\email{peterd@mpia.de}}

\date{}
 
\abstract 
{The very-high energy (VHE, $E>$100~GeV) gamma-ray sky shows diverse
Galactic and extragalactic source populations. For some sources
the astrophysical object class could not be identified so far.}
{The nature (Galactic or extragalactic) of the VHE gamma-ray source \HESSJ\ is explored.
We specifically investigate the proposed near-infrared counterpart \MASS\
of \HESSJ\ and investigate the implications of a physical association.}
{
We present K-band imaging from the 3.5 meter CAHA telescope of 2MASS~J19435624+2118233. 
Furthermore, 5 years of \textit{Fermi}-LAT data were analyzed to search for a
high-energy (HE, 100~MeV$<E<$100~GeV) counterpart.}
{The CAHA observations revealed that the near-infrared counterpart is extended with an intrinsic
half light radius of 2\arcsec\ -- 2.5\arcsec\ . These observations also show a smooth, 
centrally concentrated light profile that is typical of a galaxy, and thus point toward
an extragalactic scenario for the VHE gamma-ray source, assuming that the near-infrared source is the
counterpart of HESS J1943+213.
A high-S\'ersic index profile provides a better fit
than an exponential profile, indicating that the
surface brightness profile of \MASS\ follows that of a typical,
massive elliptical galaxy more closely than that of a disk galaxy.
With \textit{Fermi}-LAT 
a HE counterpart is found with a power law spectrum above 1~GeV, with a normalization of 
$(3.0\pm0.8_{\rm stat}\pm 0.6_{\rm
sys})\times10^{-15}\,\mathrm{cm}^{-2}\,\mathrm{s}^{-1}\,\mathrm{MeV}^{-1}$ at
the decorrelation energy $E_{\rm dec}= 15.1$ GeV and a spectral index of
$\Gamma=1.59\pm 0.19_{\rm stat}\pm 0.13_{\rm sys}$. This gamma-ray spectrum shows
a rather sharp break between the HE and VHE regimes of $\Delta \Gamma=1.47\pm0.36$.}
{The infrared and HE data strongly favor an extragalactic origin of \HESSJ\ , where
the infrared counterpart traces the host galaxy of an extreme blazar and where the
rather sharp spectral break between the HE and VHE regime indicates attenuation on
extragalactic background light. The source is most likely located at a redshift
between 0.03 and 0.45 according to extension and EBL attenuation arguments.}


\authorrunning{Peter et al.}

\titlerunning{The host galaxy and \textit{Fermi} counterpart of \HESSJ}

\maketitle


\section{Introduction}

In recent years a variety of very-high energy (VHE, $E>$100~GeV) gamma-ray sources of extragalactic
and Galactic origin have been discovered. It has been found that 
active galactic nuclei (AGN) of BL~Lac type represent the most numerous extragalactic source type 
whereas pulsar wind nebulae (PWN) are the most common Galactic VHE emitters
\citep[see][for a review]{hinton2009}.
H.E.S.S. has surveyed the region of the inner Galaxy to search systematically for VHE gamma-ray
sources \citep{aharonian2006}.

\begin{figure}[ht]
\centering
\includegraphics[height=6cm]{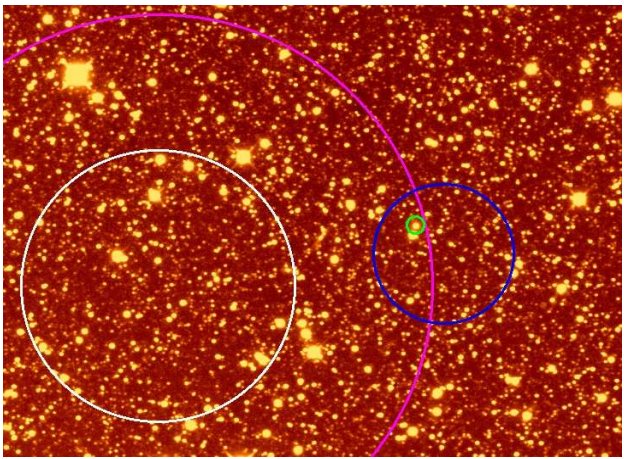}
\caption{\MASS\ and the VLBI, Chandra CXOU~J194356.2+211823, HESS J1943+213, and \textit{Fermi} objects. 
The \MASS\, VLBI and \textit{Chandra} objects lie within the 
green circle. The blue circle denotes the error (1$\sigma$) circle of the HESS observation,
the white circle is the (1$\sigma$) statistical error circle of the \textit{Fermi}-LAT observation, and the pink circle marks the (2$\sigma$) circle 
of the Fermi-LAT observation.}
\label{figure:error}
\end{figure}

During a survey of the Galactic plane H.E.S.S. discovered the point-like
source, \HESSJ\ \citep{abramowski2011}. At the position of the source
counterparts (see Figure \ref{figure:error}) in the X-ray (IGR~J19443+2117, CXOU~J194356.2+211823, SWIFT~J1943.5+2120), 
near-infrared (2MASS~J19435624+2118233) and radio regime (NVSS~J194356+211826) have been found \citep[see][]{abramowski2011}
but the nature of this source is still being debated. 
Based on
the spectral energy distribution (SED) and on the assumption that those counterparts are associated
with the VHE source, \citet{abramowski2011} interpret
this source as an extreme BL~Lac shining through the Galactic plane. In contrast
to that, \citet{gabanyi2013} found with VLBI observations that the brightness
temperature of the radio counterpart is substantially lower than expected from a
relativistically beamed BL~Lac jet and therefore argue that the object is a
young PWN. \citet{leahy2012} conclude from a HI absorption measurement that
this object is located at a distance of $>$16~kpc, which is beyond the edge of
the Galactic disk in this direction, and is thus of extragalactic origin. X-ray observations led with \textit{Suzaku} \citep{suzaku} did not reveal any variability, which would have been a smoking gun for an extragalactic origin.

The prospective near-infrared counterpart to \HESSJ\ , namely \MASS\ , 
offers a possibility to distinguish between these different interpretations and, for the purpose of this paper, we assume that \MASS\ is physically connected to
the VHE gamma-ray source.
The examination of the near-infrared counterpart is complicated by the
bright Galactic dust emission at such low Galactic latitudes ({\it b}~=~1.29$^\circ$).
For an extragalactic scenario, the host galaxy of the BL~Lac  typically would be a giant elliptical galaxy \citep[see, e.g.,][]{shaw2013}. 
Since the mean free path for VHE gamma rays from extragalactic sources is limited by 
attenuation on the extragalactic background light \citep[EBL, see][]{abramowski2013a}, any
source of such highly energetic radiation should be located in the fairly local Universe
(z $\lesssim$ 0.6). 
In case the object is a blazar, near-infrared imaging with longer integration times than what is 
provided by 2MASS \citep[Two Micron All Sky Survey,][]{cutri2003} would resolve the giant elliptical 
host galaxy and show an extended, smooth, centrally concentrated light profile.
In addition to the point-like, non-thermal infrared radiation from the AGN itself 
\citep[e.g.,][]{chen2005,massaro2011},   
the host galaxy should emit extended thermal radiation from the stellar population, 
and absorption lines in the thermal spectra would allow a
redshift measurement. For the PWN hypothesis a non-thermal spectrum would be expected 
throughout the whole nebula \citep[e.g.,][]{woltjer1958,zharikov2013}, with a faint,
diffuse near-infrared counterpart without a  clearly defined center \citep[e.g.,][]{trimble1968,fesen2008}.
To conclude, imaging and spectral observations in the infrared can help to resolve the controversy
over the nature of \HESSJ.

In addition to examining the near-infrared counterpart, exploring any high energy (HE, 100~MeV$<E<$100~GeV)
counterpart
will also lead to important constraints on the nature and distance of the VHE emitter. The currently
operating Large Area Telescope (LAT) onboard the \textit{Fermi} satellite provides deep observations of \HESSJ\ in this energy range.
Since VHE emission of distant extragalactic objects is attenuated by EBL 
whereas the HE radiation is much less affected by this attenuation, the connection of the HE and VHE
spectrum can give evidence of an extragalactic origin of a gamma-ray emitter \citep{abdo2010,sanchez2013}. 

This paper is structured in the following way. In Sec.~\ref{sec:pop} all infrared sources within the
positional error circle of \HESSJ\ are investigated, in Sec. \ref{sec:caha} deep K-band observations of
the most likely infrared counterpart are presented, in Sec. \ref{sec:fermi} the results of the
LAT data analysis are given and in Sec. \ref{sec:disc} constraints on the distance to the object and 
SED modeling of the gamma-ray source are discussed. Throughout this paper a $\Lambda$CDM cosmology with 
H$_0$~=71~km~s$^{-1}$~Mpc$^{-1}$, $\Omega_\Lambda$ = 0.73, and $\Omega_\mathrm{M}$ = 0.27 is assumed.

\section{Infrared sources inside the error circle of the VHE gamma-ray source}\label{sec:pop}

In this section, color-color diagrams are used to check
whether the properties of the prospective infrared counterpart to \HESSJ, namely \MASS, 
are consistent
with other sources in the vicinity or if it represents an
outlier with respect to the general properties of infrared sources there.
The spectra of extragalactic sources are affected by the absorption by interstellar
dust and are expected to be redder than foreground objects. Furthermore,
non-thermal sources may occupy different regions in color-color diagrams
with respect to thermal sources.
The data for this investigation were retrieved from
CDS Strasbourg through the Aladin tool \citep{bonnarel2000} and further processed 
with TOPCAT \citep{taylor2005}. The region around \HESSJ\ was observed by 
the 2MASS, the UKIDSS \citep[UKIRT Infrared Deep Sky Survey,][]{warren2007}, and the WISE \citep[Wide Field Infrared Survey Explorer,][]{wright2010} surveys.
\MASS\ was detected by each of the surveys. 2MASS detects it only in the K-band 
and gives upper limits for J and H, while UKIDSS 
gives detections in J, H and K. Therefore, we took only the UKIDSS data of \MASS\ 
for further evaluation. The UKIDSS catalogue entries of \MASS\ are as follows: 
$\alpha_{\rm J2000}=19^{\rm h}\,43^{\rm
m}\,56.24^{\rm s}$, $\delta_{\rm J2000}=21^\circ\,18'\,23.3\arcsec$, ID J194356.23+211823.3, 
${\rm J}=16.448 \pm 0.010$~mag,   
${\rm H}=15.187 \pm 0.006$~mag,  
${\rm K}=14.174 \pm 0.006$~mag, and based on its spatial extent it has been classified as a galaxy with a 10\%
probability of being a star.

\begin{figure}[ht]
\centering
\includegraphics[height=6cm]{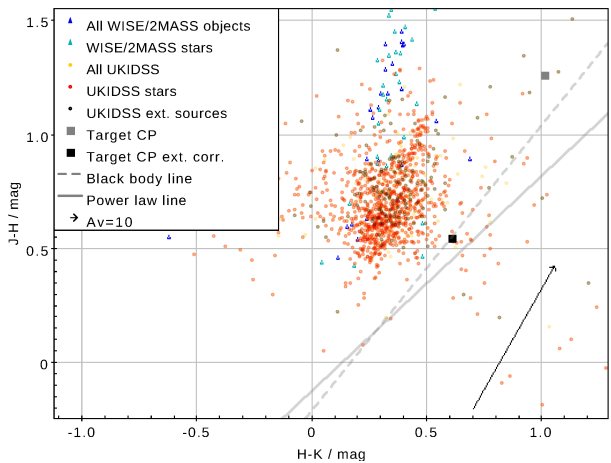}
\caption{Color-color plot in the J, H and K bands for the infrared sources within the error circle of \HESSJ.
The infrared counterpart \MASS\ is clearly redder than the main distribution of sources in this direction of the sky.
The arrow indicates the direction of reddening by extinction by interstellar dust for A$_\mathrm{V}$ = 10.}
\label{figure:jhk}
\end{figure}

All objects within a radius of 1.5\arcmin\ around the source coordinate, $\alpha_{\rm J2000}= 19^{\rm h}\,43^{\rm
m}\,56.23^{\rm s}$, $\delta_{\rm J2000}=21^\circ\,18'\,23.3\arcsec$, were used to construct a color-color diagrams. The radius of the region
was chosen to coincide with the 99\% positional confidence level of the H.E.S.S. source
as shown in \citet{abramowski2011}.\newline
Figure \ref{figure:jhk} shows the J-H-K
diagram for all the objects retrieved from the database. The 2MASS and UKIDSS data
show a concordant main distribution. 
The proposed infrared counterpart to \HESSJ\ is outside of the main
concentration of sources. The object is redder than 
most sources in H$-$K as well as in J$-$H. 
The data points were not corrected for extinction by dust in the Galactic disk
since the positions of the sources inside the Galaxy and thus actual extinctions are in most cases not known.
Even after applying the extinction correction only
for the counterpart of \HESSJ\ (for extragalactic objects outside the Galactic disk
the entire Galactic extinction applies) the object still
remains off the main distribution. \newline
To quantify the probability that the \HESSJ\ belongs to the general distribution of objects in the error circle of 1.5\arcmin\ of
the HESS telescope we constructed a two-dimensional histogram of the UKIDSS color-color data. A Gaussian fit was applied to the histogram.
This Gaussian is tilted by $\phi$ = -24 degrees with respect to the axis of the color-color diagram.
It has a minor axis standard deviation of $\sigma_1$ = 0.09 mag and a  major axis standard deviation of $\sigma_2$ = 0.16 mag. 
The distance between \HESSJ\ and the 
center of the Gaussian is 7.8 $\sigma$ with
\begin{equation} 
\sigma = \left |\left ( \begin{array}{c c} cos(\phi)& sin(\phi)\\ -sin(\phi)& cos(\phi) \end{array}\right )
\left(\begin{array}{c}\sigma_1 \\\sigma_2\end{array}\right )\right |.
\end{equation}
 As the extinction is not well known in this region of the Galaxy we 
checked the closest distance between a line through the position  of \HESSJ\ and the extinction direction 
(A$_\mathrm{V}$ arrow) \citep[extinction correction
follows][i.e. A$_\mathrm{J}$ = 1.952 mag, A$_\mathrm{H}$ = 1.936 mag, A$_\mathrm{K}$ = 0.832 mag]{schlafly2011} in the color-color 
diagram. 
This closest distance is still 2.4 $\sigma$. This position is reached for a differential extinction between the main distribution and 
 \HESSJ\ of about 10 mag.\newline
The parameters of the Gaussian fit are the same within $5\%$ for histograms with different binnings, and there are
minor changes to the analysis above if one only uses the distribution of stars only instead of the distribution
of all objects.\newline
  Figure \ref{figure:jhk} shows also the black body line (for redshift z = 0) 
and the power law lines, which were
taken from \citet{chen2006}. These lines were obtained with the measured value for the Mauna Kea
Observatories Near-Infrared Filter Set
\citep{tokunaga2005}, which are used for the UKIDSS survey (J = 1.25 $\mu$m, H = 1.644 $\mu$m,
K$_\mathrm{s}$ = 2.198 $\mu$m, S$_\mathrm{J}$ = 1560 Jy, S$_\mathrm{H}$ = 1040 Jy,
S$_\mathrm{K}$ = 645 Jy).
The extinction corrected measurement of \MASS\ falls on the black body line,
which would lead to the conclusion 
that the object is thermally dominated\footnote{This thermal emission is related to the host galaxy, see section \ref{sec:disc}.} in the near-infrared.
However, due to the large and potentially uncertain extinction, this conclusion has to be taken with caution.

\begin{figure}[ht]
\centering
\includegraphics[height=6cm]{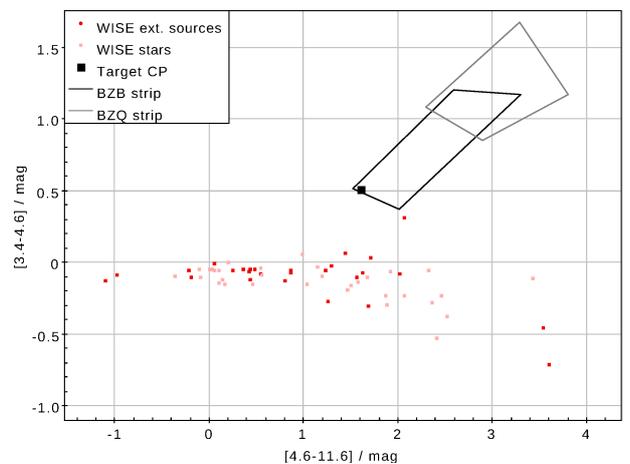}
\caption{Color-color plot in the WISE W1, W2 and W3 bands for the infrared sources within the \HESSJ\ error circle. 
\MASS\ is located within the so-called `Blazar strip'. No correction for extinction on interstellar dust has been applied. Note that BZB stand for BL Lac objects while BZQ for flat-spectrum radio quasars.}
\label{figure:wise}
\end{figure}

In a second step we studied the mid-infrared measurements of the WISE survey. 
WISE observed the sky in four mid-infrared bands: W1 = 3.4 $\mu$m, W2 = 4.6 $\mu$m,
W3 = 12 $\mu$m, and W4 = 22 $\mu$m.
The AllWISE catalogue entry for the counterpart to \MASS\ is: ID~J194356.24+211822.9,
position: $\alpha_{\rm J2000}= 19^{\rm h}\,43^{\rm
m}\,56.24^{\rm s}$, $\delta_{\rm J2000}=21^\circ\,18'\,23.3\arcsec$,
W1 = $13.056 \pm 0.116$~mag,
W2 = $12.523 \pm 0.094$~mag, W3 = $10.909 \pm 0.180$~mag,
W4 $> 8.144$~mag.
In Figure \ref{figure:wise} the [3.4]-[4.6]-[12] $\mu$m color-color diagram is
shown. No correction for extinction on interstellar dust has been applied.
However, it has to be noted that mid-infrared radiation is only mildly affected
by dust extinction. The empirically found blazar strips of \citet{massaro2011}
are also drawn in this diagram. 
These strips represent the area in this color-color diagram where blazars lie. This is interpreted as due to their mid-infrared colors being consistent with non-thermal radiation.
\MASS\ is clearly located at the edge of the
Blazar strip in the mid-infrared, indicating that the emission in this energy
band is dominated by non-thermal emission.

To summarize, \MASS\ appears to be of extragalactic origin and can likely be classified
as a blazar.

\section{CAHA observations}\label{sec:caha}

On July, 25 and 26, 2013 we observed the region around \HESSJ\ with the wide field camera OMEGA2000 installed on the 3.5~m telescope at 
Calar Alto \citep{kovacs2004} of the Centro Astronomico Hispano Aleman. 
On July, 25 we used the K-short filter and on July, 26 K-prime.
The field was centered on the X-ray counterpart 
of the object. The target was observed 30 min in total each night. For sky subtraction we took separate sky frames 
as the object is in the Galactic plane.
The seeing was 1.1\arcsec\ on the first and 1.6\arcsec\ on the second night with poor transparency during both nights. 

\begin{figure}[ht]
\centering
\includegraphics[height=8cm]{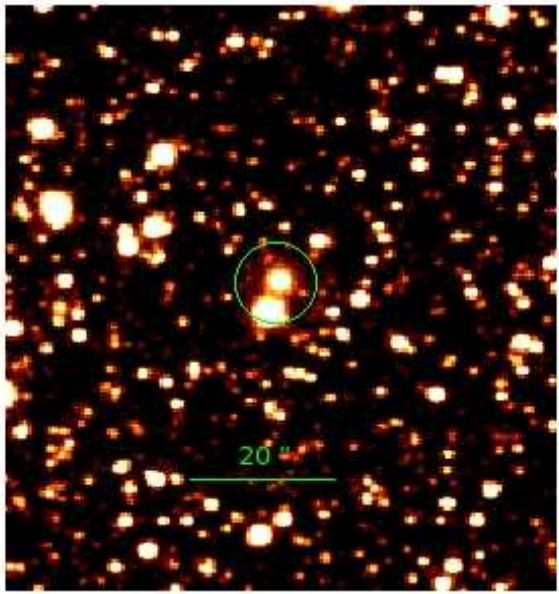}
\caption{The object and its neighbourhood is shown. The green circle is centered on the target. The object in the lower part 
of the circle is one of the stars ($\alpha_{\rm J2000}= 19^{\rm h}\,44^{\rm
m}\,00.0^{\rm s}$, $\delta_{\rm J2000}=21^\circ\,17'\,16.1\arcsec$) used to determine the target coordinates.}
\label{figure:Region}
\end{figure}

Fig. \ref{figure:Region} shows the near-infrared counterpart and its neighborhood.
The position of the object was determined using 8 bright stars of the 2MASS catalogue within a circle of
$3'$ radius. The coordinate system data were scaled and rotated using the 28 connecting lines between these stars.
The position of \MASS\ was determined to be 
$\alpha_{\rm J2000}= 19^{\rm h}\,43^{\rm
m}\,56.3^{\rm s} \pm 0.1^{\rm s}$, $\delta_{\rm J2000}=21^\circ\,18'\,23.4\arcsec \pm 0.2\arcsec$
which agrees with the VLBI position to better than $0.1\arcsec$ \citep{gabanyi2013}. 
The flux of the object was determined to be 13.5 mag $\pm$ 0.3 mag in K.
The higher brightness with respect to the UKIDSS measurement results
from faint outer regions of the extended source that were missed by the shallower UKIDSS image.
We checked the variation of the flux between single exposures and night 1 and night 2.
The maximum variation in the flux was measured to be 0.5 mag, which is of the level of measured flux variations 
due to sub pixel shifts of the object. Thus no variability in the timespan of 30 min and between night 1 and night 2 
was detected.

To classify the object and its morphology we used two different approaches. With the first approach we used {\tt{galfit}} \citep{peng2010}.
The object was modeled with a S\'ersic profile \citep{sersic1963}:
\begin{equation}I[r]=I_e\exp(-a(r/r_e)^{1/n}-1)\end{equation}
with $I[r]$ the intensity at radius $r$, $I_e = I[0]/e$, $a$ a normalization factor, $r_\mathrm{e}$ the half light radius, and $n$ the S\'ersic parameter.

\begin{figure*}
\centering
\includegraphics[width=0.45\textwidth]{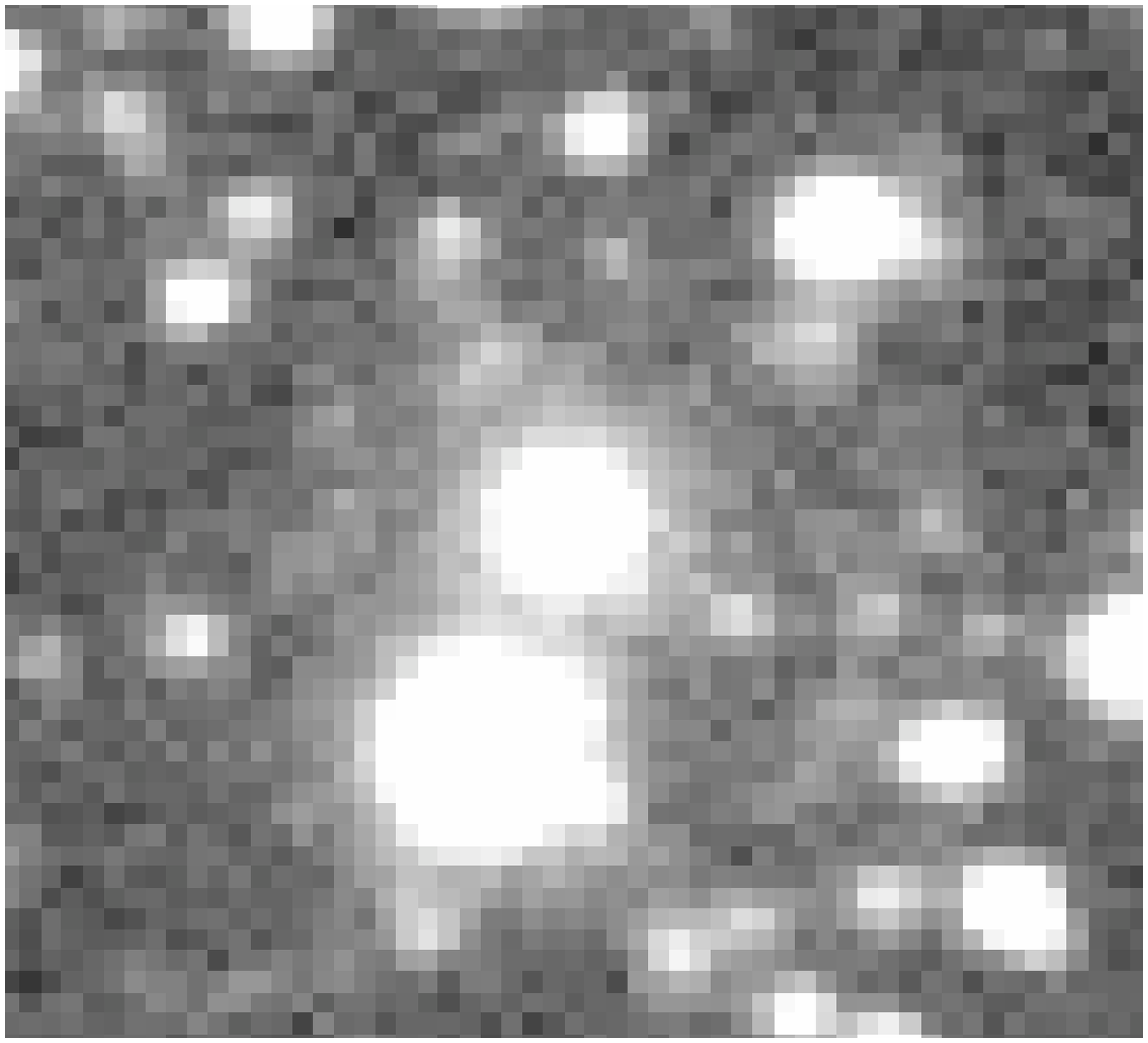}
\includegraphics[width=0.45\textwidth]{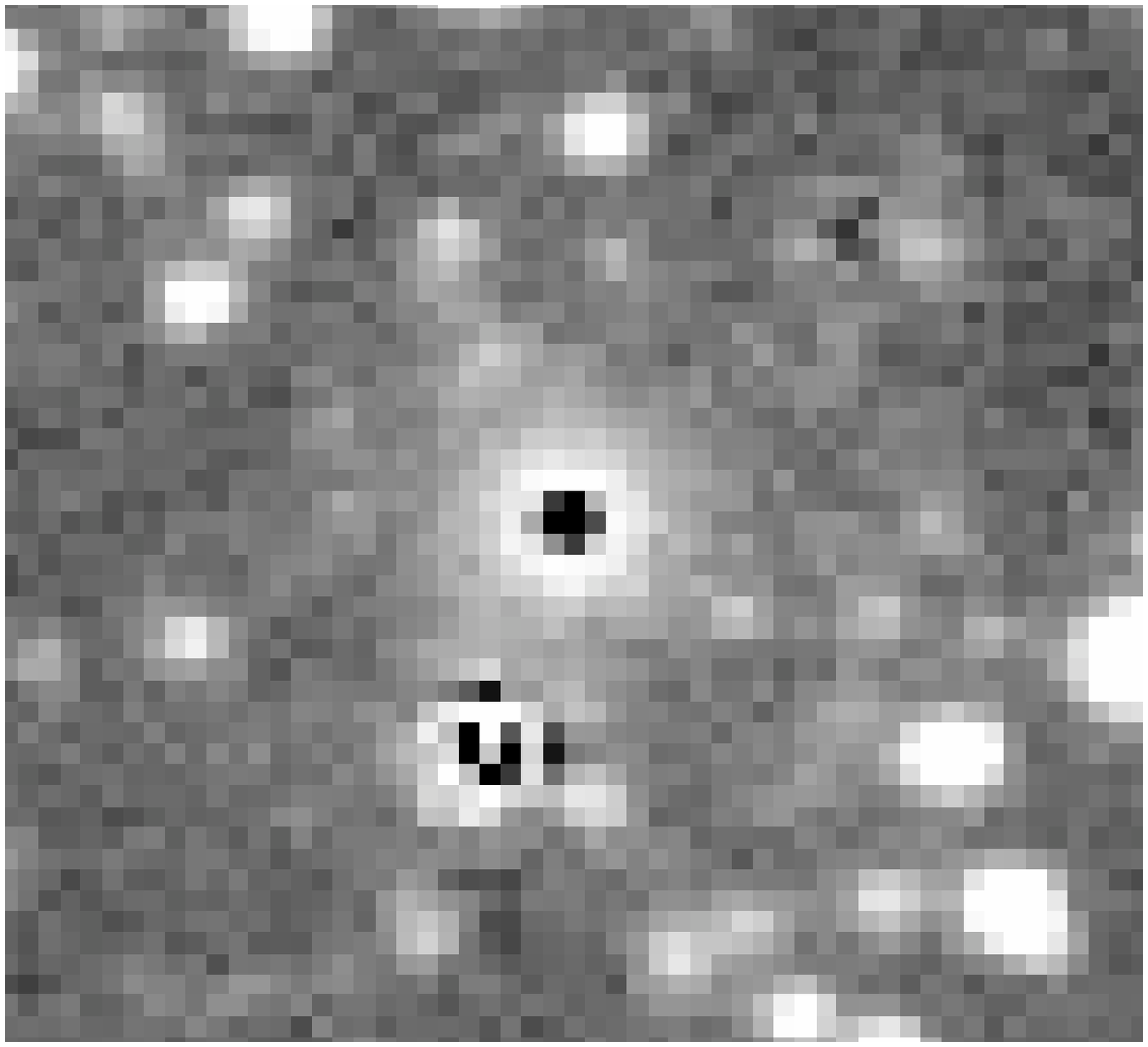}
\includegraphics[width=0.45\textwidth]{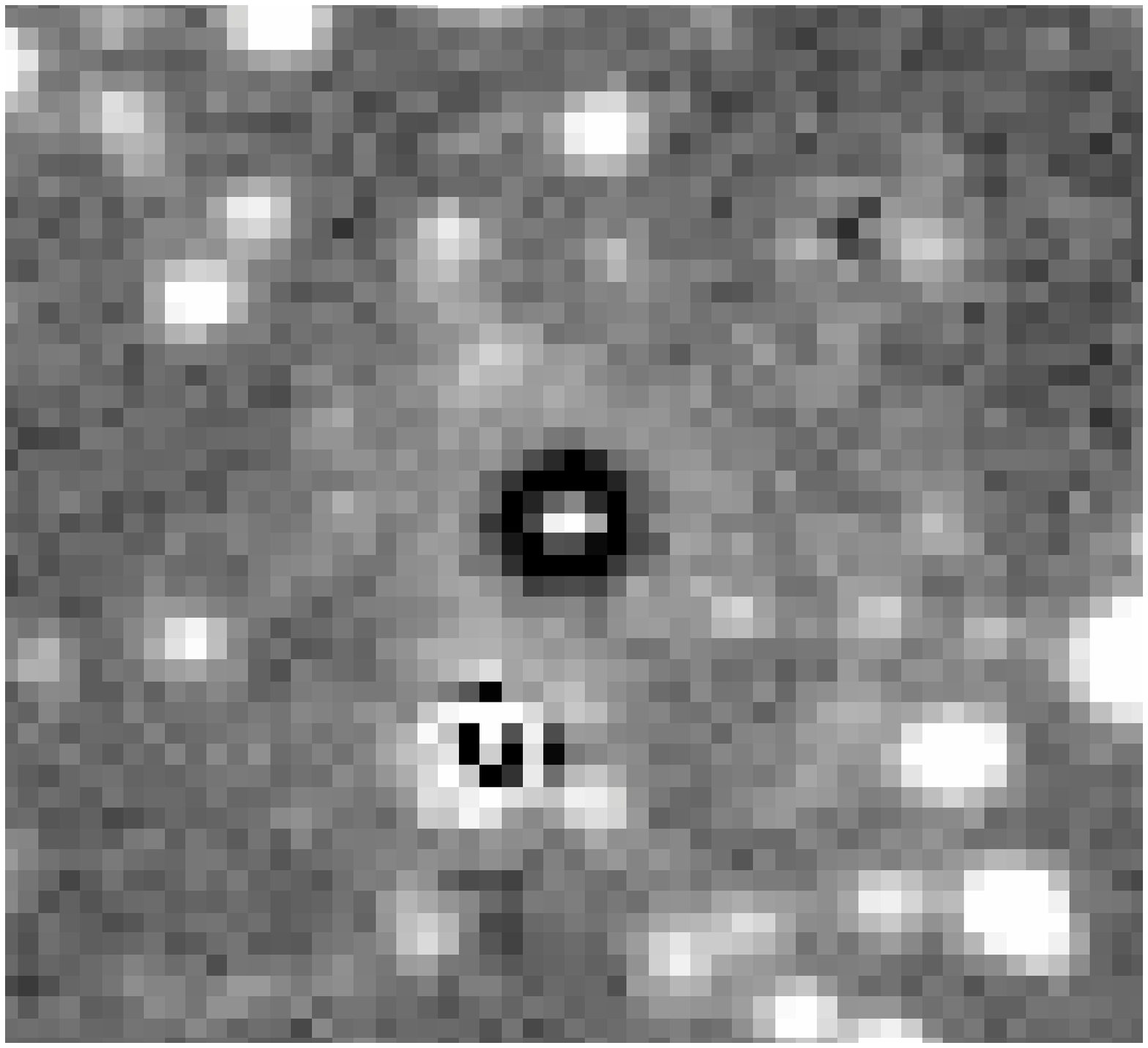}
\includegraphics[width=0.45\textwidth]{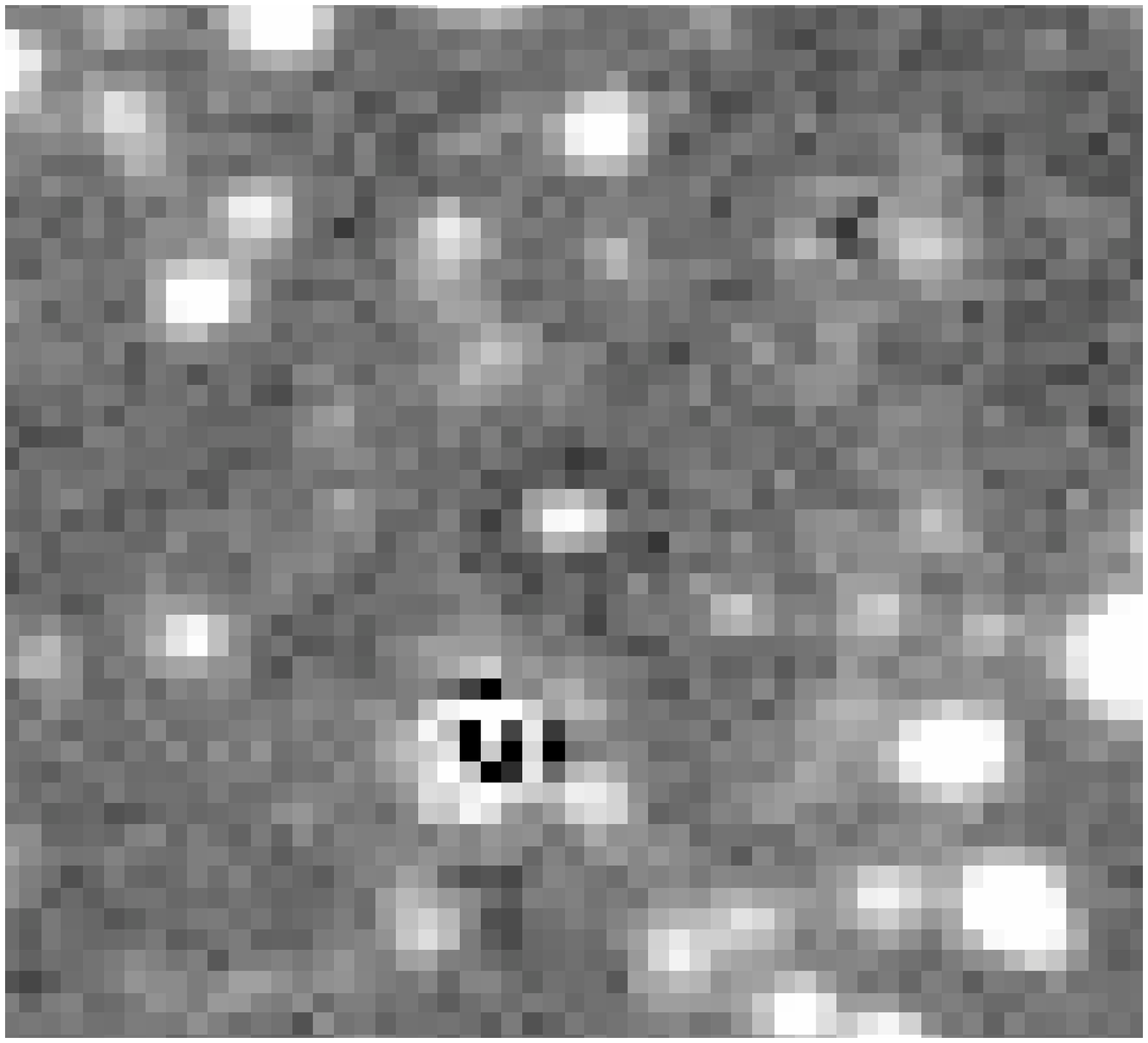}
\caption{K-band image of the infrared counterpart of \HESSJ\ (25\arcsec\ $\times$ 25\arcsec\ ).
The image \textbf{at the top left} is centered on \MASS\, the prospective infrared counterpart
of the VHE gamma-ray source. The next 3 panels show
residuals after subtracting, at the top \textbf{right}, a point source, 
at the bottom row from left to right, an
exponential surface brightness profile, and a surface brightness
profile with S\'ersic index 7. The package {\tt galfit } was used to determine the
best-fitting free parameters (magnitude and, for the extended
profiles, half-light radius). Bright, neighboring stars were also
subtracted to avoid confusion. The prominent residual after
subtracting a point source clearly demonstrates that the infrared counterpart
to \HESSJ\ is
extended (to be compared with the residual from the bright star directly
below \HESSJ). Furthermore, a high-S\'ersic index profile leaves a
smaller residual than an exponential profile, indicating that the
surface brightness profile of \HESSJ\ follows that of a typical,
massive elliptical galaxy more closely than that of a disk galaxy. The
moderate residual for the high-S\'ersic index model hints at the presence of
a central point source, but imaging with higher resolution is needed to confirm this.
}
\label{fig:morph} 
\end{figure*}

Due to the number of sources in the field we could not let the program find the S\'ersic index and half light radius
automatically but had to rerun the program for different values of these parameters to find the best fit.
In Fig. \ref{fig:morph} the morphology of the source is investigated.
Subtracting a point-like source clearly leaves residuals, thus showing that the object is extended.
These residuals are significant at the 50$\sigma$ confidence level when compared to background fluctuations
in blank fields of the sky.
In order to determine if the object could be  a disk galaxy or an elliptical we made a best-fit approach with an exponential
profile and S\'ersic index profiles with different indices.
We found that a high-S\'ersic index profile leaves a
smaller residual than an exponential profile, indicating that the
surface brightness profile of \HESSJ\ follows that of a typical,
massive elliptical galaxy more closely than that of a disk galaxy.
The object is round (axis ratio $>$0.97) and has a half light radius of $\approx$ 2.0\arcsec\ $<r_\mathrm{e}<$ 2.5\arcsec .
The S\'ersic index $n$ for the best fit with {\tt{galfit}} was determined to be 8.\newline
We double checked the results with a custom algorithm in {\tt IDL}. The program was used to fit a point-spread function (PSF) taken from the image of a neighboring star and a S\'ersic profile convolved with the PSF
to the data. 
Here we left the positions and half light radius free to vary but fixed the S\'ersic parameter.
For the PSF fitting we used different stars. The alternative method also yielded a half light radius of $\approx$ 2.0\arcsec $<r_\mathrm{e}<$ 2.5\arcsec.
 The S\'ersic index $n$ for the best fit was determined to be $>$ 7.
Therefore, the results from the two methods are consistent. A moderate residual for the high-S\'ersic index model suggests 
the presence of a central point source. Higher-resolution imaging will be needed to confirm this feature.

To summarize, \MASS\ appears to be an elliptical galaxy and can likely be classified
as a blazar. Given the correspondence of the position between \MASS\ and the VLBI and \textit{Chandra} object 
it can be assumed that it is the
same object. Given this multi-wavelength property
of the object we conclude that \MASS\ is a viable counterpart to HESS~J1943+213.

\section{\textit{Fermi}-LAT observations}\label{sec:fermi}

The LAT is sensitive to gamma rays from 20
MeV to $>$ 300 GeV. A detailed description of the instrument and its performance is
given in \citet{atwood2009}. The LAT mainly operates in survey mode in which the entire sky is observed every three hours.

For the analysis of the LAT data, the events belonging to the class {\tt SOURCE}
\citep{ackermann2012} have been retained. Considering that the object is
located in the Galactic plane, the energy range has been restricted to 1-300~GeV.
This cut reduces the diffuse Galactic background which presents a
rather soft spectrum compared to a hypothetical LAT counterpart of a TeV
blazar. A region of interest of $10^\circ$ radius around the coordinates of
\HESSJ\ was defined to perform a binned analysis \citep{mattox1996}, implemented in
the {\tt gtlike} tool, part of the ScienceTools\footnote{\url{http://fermi.gsfc.nasa.gov/ssc/data/analysis/software/}} {\tt V9R32P5}. Cuts were applied
on the rocking angle of the spacecraft, which was required to be smaller than
$52^\circ$, and on the zenith angle of the events, required to be smaller than
$100^\circ$ to reduce Earth limb gamma-rays. The {\tt Pass 7} Reprocessed data
\citep{bregeon2013}, from August, 4 2008 to August, 4 2013, were
used in this study together with the corresponding instrument response
functions {\tt P7REP\_SOURCE\_V15}. The last Galactic diffuse model ({\tt gll\_iem\_v05\_rev1.fit}) and the public isotropic  model ({\tt iso\_source\_v05.txt}) were included in the sky model\footnote{Files are available at \url{http://fermi.gsfc.nasa.gov/ssc/data/access/lat/BackgroundModels.html}.} as well as all the sources within
$15^\circ$ around \HESSJ\ present in a  preliminary list of sources, currently only available to the LAT Collaboration, based on 4 years of data (3FGL). Sources within $3^\circ$ of the target have their spectral parameters free to vary during the optimization procedure and the other parameters are frozen to the best fit values of the 4-year list. Finally a source, with a power law spectrum, was
added to the sky model at the position of \HESSJ. After an initial fit, the sources with Test 
Statistic\footnote{see \url{http://fermi.gsfc.nasa.gov/ssc/data/analysis/documentation/Cicerone/Cicerone_Likelihood/Likelihood_overview.html} for a definition.} lower than 1 were removed and the fitting procedure was redone.

A counterpart to \HESSJ\ is detected above 1 GeV. To better locate the GeV
counterpart, the tool {\tt gtfindsrc} has been used yielding a position
$\alpha_{\rm J2000}= 19^{\rm h}\,44^{\rm m}\,05^{\rm s} $, $\delta_{\rm
J2000}=21^\circ\,17'\,51\arcsec$ with a $1\sigma$ error circle radius of
$1'\,12\arcsec$ (error is statistical only), at an angular separation of $2'\,24\arcsec$ 
from the position of HESS J1943+213. At this location, the counterpart is
detected above 1 GeV at a TS of 36.0 (corresponding to $\approx5.1\,\sigma$ for
4 degrees of freedom\footnote{The degrees of freedom are two spectral parameters
and the position (two parameters).}). Taken the LAT systematical uncertainties
\citep{2012ApJS..199...31N} and H.E.S.S. uncertainties\footnote{The best
position reported is $\alpha_{\rm J2000}=19^{\rm h} 43^{\rm m} 55^{\rm s} \pm
1^{\rm s}_{\rm stat} \pm 1^{\rm s}_{\rm sys}$, $\delta_{\rm J2000}= +21^{\circ}
18' 8'' \pm 17''_{\rm stat} \pm 20''_{\rm sys}$. }, the positions are marginally
in agreement (below $2\sigma$). The good agreement of the HE and VHE spectra
(see later and Figure~\ref{figure:sed}) also favor a common origin of the
measured emission.

 Using the LAT position, the resulting best-fit spectrum is a power law with a normalization of 
$(3.0\pm0.8_{\rm stat}\pm 0.6_{\rm
sys})\times10^{-15}\,\mathrm{cm}^{-2}\,\mathrm{s}^{-1}\,\mathrm{MeV}^{-1}$ at
the decorrelation energy $E_{\rm dec}= 15.1$ GeV and a spectral index of
$\Gamma=1.59\pm 0.19_{\rm stat}\pm 0.13_{\rm sys}$ yielding an integral flux between 1~GeV and 300~GeV of $I=(3.7\pm 1.3_{\rm stat}\pm 0.8_{\rm
sys})\times10^{-10}\,\mathrm{cm}^{-2}\,\mathrm{s}^{-1}$. The highest energy photon that can be associated with the counterpart (with a probability higher than 80\%) has $E=51$\,GeV. Systematic uncertainties from the modeling of the LAT effective area were evaluated using the IRF bracketing method \citep{abdo2009} and by using alternative interstellar emission models \citep{2013arXiv1304.1395D} for the Galactic diffuse model. Data points were obtained by repeating the analysis in restricted energy ranges equally spaced in log energy.
No variability has been found for this source using an unbinned Bayesian block analysis \citep{scargle2013}. Note that, since this source is in the Galactic plane, the modeling of the Galactic diffuse emission might have a significant impact on the spectral parameters which has been evaluated and taken into account in the systematical uncertainties. \citet{suzaku} did not find a significant detection of the source above 10 GeV using 4.5 years of Pass 7 data. They reported a low-significance spectral measurement in the range 10~GeV-300~GeV with an index of $\Gamma\approx2.4$. Repeating the analysis described here in the same energy range leads to an index of $\Gamma=2.26_{-0.47}^{+0.53} $. This can be interpreted as the sign of a rollover of the photon spectrum and is in agreement with the softer VHE measurement of H.E.S.S.

\section{Discussion}
\label{sec:disc}
\subsection{Host galaxy and redshift}

In this paper we have studied \MASS, which we assume to be the infrared counterpart 
of \HESSJ\ and additionally,
we have detected the source in HE gamma-rays with the LAT. These
observations place important constraints on the nature of the
object. 
We investigated all infrared sources within the error circle
of \HESSJ\ and found that \MASS\ is redder than the bulk of infrared sources at this location on the sky
and that it falls in the blazar strip \citep{massaro2011} in the mid-infrared. 
Using the 3.5 m CAHA telescope we found that the K-band counterpart \MASS\ is extended.
These
observational results strongly suggest that \MASS\ is of extragalactic origin.
The extended counterpart of the 2MASS source represents the host galaxy of a BL~Lac type object and is a viable counterpart to HESS~J1943+213.
The extension of the
object can even give constraints on the distance to the host galaxy.
BL Lac host galaxies typically have infrared half radii of $3.1 \pm 1.7$~kpc
\citep{cheung2003}\footnote{scaled to the adopted cosmology}. For the measured
half light radius of \MASS\ of 2.5\arcsec\ this would correspond to
an angular diameter distance $D_\mathrm{A}$ and redshift $z$ of $D_\mathrm{A} = 120$~Mpc ($z \approx 0.03$) for a half light radius of 1.4~kpc, 
$D_\mathrm{A} = 260$~Mpc ($z\approx 0.07$) for a half light radius of 3.1~kpc and $D_\mathrm{A} = 410$~Mpc ($z \approx 0.11$) 
for a half light radius of 4.8~kpc \citep{wright2006}. 
Since the half-light radius
of the BL Lac host is not very well defined, this places only loose
constraints on the distance and redshift of the object. The lower limit adopted in this work is $z>0.03$.
For comparison, using the typical K-band luminosity of BL Lac host galaxies
\citet{abramowski2011} found a redshift of $z \gtrapprox 0.14$.

\begin{figure}[ht]
\centering
\includegraphics[height=6cm]{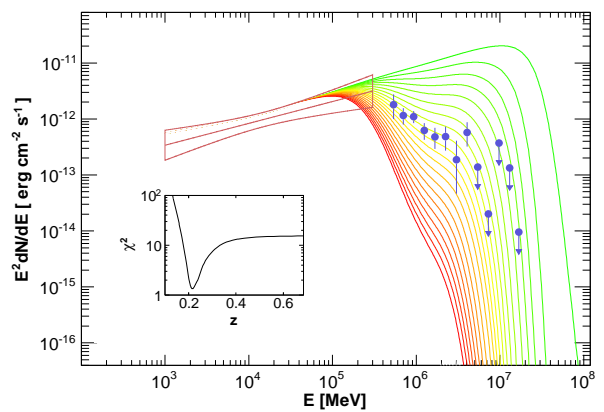}
\caption{Spectral energy distribution of \HESSJ\ in $\gamma$-rays. 
The \textit{Fermi}-LAT error contour (red) has been extrapolated towards 
the H.E.S.S. energy range (blue points) and corrected for EBL absorption 
using the model of \citet{franceschini2008} assuming different values for 
$z$ from $z=0.01$ (green line) to $z=0.7$ (red line). The inset gives the corresponding $\chi^2$ as a function of the redshift $z$.}
\label{figure:z}
\end{figure}

A constraining upper limit on the redshift can be obtained by comparing the H.E.S.S. data points with the 
\textit{Fermi}-LAT results, extrapolated to the TeV energies 
and corrected for EBL absorption assuming different values for $z$. 
The model of EBL used in this work is described in  \citet{franceschini2008}. For each extrapolation, 
a $\chi^2$ is then computed to estimate the most likely value of $z$ \citep{abdo2010}. To be conservative, the error on differential flux  computed using equation~2 of \citet{abdo2010} was added to the differential flux. Figure~\ref{figure:z} gives the results for different values of $z$ as well as the evolution of the 
$\chi^2$ as a function of the redshift. A minimum is reached for $z\approx 0.22$. An upper limit at 0.45 can be set with a confidence level of 95\%.

Another method was proposed by \citet{sanchez2013} to roughly estimate the
redshift based on the spectral break between the GeV and TeV energy ranges.
Applying this method leads to an upper limit of $z<0.39$, which provides a
cross-check with the previous method. Summarizing, if \HESSJ\ is assumed to be \MASS\
our findings limit its redshift to be between 0.03 and 0.45.

\subsection{SED modeling}\label{sec:sed}

Blazars present a double humped spectral energy distribution (SED) with a
low-energy (from radio to X-ray) peak attributed to synchrotron emission of
relativistic electrons. The origin of the high-energy component is still uncertain.
The leptonic class of models invoke the production of gamma-rays by inverse
Compton scattering on synchrotron photons \citep[synchrotron self-Compton,
SSC,][]{band1985} or on an external field \citep[external Compton,][]{dermer1993}

\begin{figure}[ht]
\centering
\includegraphics[height=6cm]{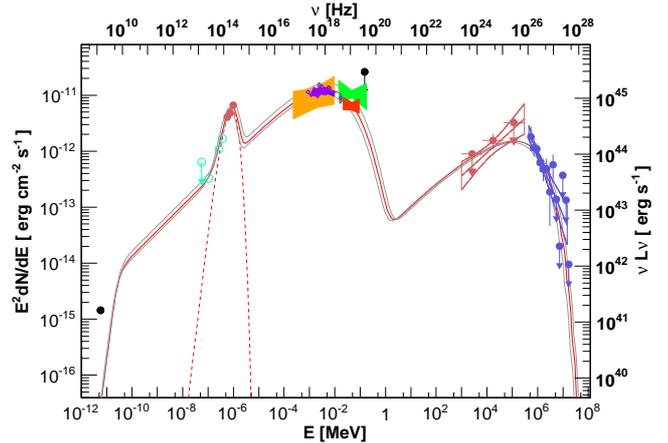}
\caption{Spectral energy distribution of \HESSJ. Empty blue circles are the WISE data and the 
red circles the CAHA results. The red butterfly is the \textit{Fermi}-LAT best-fit power law and the 
68\% error contour. A 2-$\sigma$ upper limit is reported for the LAT data points if TS $<$ 9. Other data were extracted from \citet{abramowski2011} : The black point is the NVSS measurement, the yellow, blue, and red butterflies are respectively the \textit{Chandra}, \textit{Swift}-XRT, and INTEGRAL-IBIS measurements. The green point is from the 70-month \textit{Swift}-BAT catalogue \citep{2013ApJS..207...19B}. The upper limit is from INTEGRAL-SPI. The purple points are the \textit{Suzaku} measurements extracted from \citet{suzaku}. 
The red line is the result of the SSC calculation for $z=0.2$, and the grey lines were obtained for $z=0.03$ and $z=0.45$.}
\label{figure:sed}
\end{figure}

An SSC model has been used to reproduce the SED of the source presented in
Figure~\ref{figure:sed}. A single spherical  emission zone of size $R$ filled with
a uniform magnetic field $B$ has been considered. The emission zone is moving
relativistically towards the Earth with a Doppler factor $\delta$. The electron
distribution, $N_e(\gamma)$, is described by a power law with an exponential
cut-off of the form $N_e(\gamma)\propto\gamma^{-p}\cdot\exp(-\gamma/\gamma_{\rm
cut})$ from $\gamma_{\rm min}=1$ to $\gamma_{\rm max}=10^{10}$. A redshift of
0.2 has been assumed for the SSC calculation, which leads to a luminosity distance
of $D_L= 971$~Mpc \citep{wright2006}. The results of the SSC calculation
and the black-body emission are shown in Figure~\ref{figure:sed} and the
parameters are given in Table \ref{table:ssc}. For completeness, the
calculations for $z=0.03$ and $z=0.45$ are shown in Figure~\ref{figure:sed}.

The host galaxy is fitted using a black-body model \citep{katarzynski2003} assumed to 
originate from thermal emission, with a
temperature of 3000 K and a total luminosity of $4.5\times 10^{44}$~erg\,s$^{-1}$.
The near-infrared measurements rule out a pure power law for this energy regime.
In this calculation, the jet is out of equipartition, the ratio between the kinetic energy 
of the electrons and the magnetic energy being $Q = u_e/u_B=12$. Exhibiting a jet dominated 
by the energy of the electrons seems to be a feature of the TeV  high-frequency peaked BL~Lacs, see e.g. Mrk~421
\citep{abdo2011a}, Mrk~501 \citep{abdo2011b}, SHBL~J001355.9$-$185406 
\citep{abramowski2013b} or 1ES~1312$-$423 \citep{abramowski2013c}. 

\begin{table}
\caption{Parameters values of the SSC model used to reproduce the SED of \HESSJ.}
\label{table:ssc} 
\centering 
\begin{tabular}{c  c } 
\hline\hline 
Parameters & Value     \\ 
\hline 
$B$ [G] & $0.05$  \\
$R$ [cm] & $8.5\times 10^{16}$   \\
$\delta$ & $10$  \\
\hline 
$p$ & $2.1$  \\
$\gamma_{\rm cut}$ & $10^6$\\
$N_{\rm tot}$ & $4.7\times 10^{53}$ \\
\hline 
\end{tabular}
\end{table}

\section{Outlook}

The nature of the VHE emitter \HESSJ\ has been disputed in the past
\citep{abramowski2011,gabanyi2013,leahy2012} with diverging conclusions
whether the object is of Galactic or extragalactic origin.
In this work we have found that the infrared and HE counterparts show a consistent picture that points
towards an extreme blazar hosted by a giant elliptical galaxy. Additionally, the infrared surface brightness 
of the prospective host galaxy appears to be more centrally peaked than expected for 
a giant elliptical galaxy, thus indicating that a central point source may be present in this system. 
Further observations in radio and possiblily infrared will be required
to finally settle the unexplained properties of this enigmatic source.

\begin{acknowledgements}

DS work is partially supported by the LABEX grant enigmass.
The authors want to thank T. Cheung for discussions
on the host galaxies of BL~Lac's and an anonymous referee for helpful comments. D. P., W. D. and W. G.
want to thank the Pinte for its inspiring spirits and the
EMBL cantine for the sustenance, which supported us during all this work.

The \textit{Fermi} LAT Collaboration acknowledges generous ongoing support
from a number of agencies and institutes that have supported both the
development and the operation of the LAT as well as scientific data analysis.
These include the National Aeronautics and Space Administration and the
Department of Energy in the United States, the Commissariat \`a l'Energie Atomique
and the Centre National de la Recherche Scientifique / Institut National de Physique
Nucl\'eaire et de Physique des Particules in France, the Agenzia Spaziale Italiana
and the Istituto Nazionale di Fisica Nucleare in Italy, the Ministry of Education,
Culture, Sports, Science and Technology (MEXT), High Energy Accelerator Research
Organization (KEK) and Japan Aerospace Exploration Agency (JAXA) in Japan, and
the K.~A.~Wallenberg Foundation, the Swedish Research Council and the
Swedish National Space Board in Sweden.
 
Additional support for science analysis during the operations phase is gratefully
acknowledged from the Istituto Nazionale di Astrofisica in Italy and the Centre National d'\'Etudes Spatiales in France.

\end{acknowledgements}

\end{document}